\documentstyle[11pt,aaspp4]{article}

\lefthead{}
\righthead{Infrared Spectroscopy of $\epsilon$ Indi B}

\begin{document}

\title{High-Resolution Infrared Spectroscopy of\\
the Brown Dwarf $\epsilon$ 
Indi Ba \footnote{Based on observations
obtained at the Gemini Observatory, which is operated by the Association
of Universities for Research in Astronomy, Inc., under a cooperative
agreement with the NSF on behalf of the Gemini partnership: the National
Science Foundation (United States), the Particle Physics and Astronomy
Research Council (United Kingdom), the National Research Council (Canada),
CONICYT (Chile), the Australian Research Council (Australia),
CNPq (Brazil), and CONICRT (Argentina).}$^{,}$ \footnote{Based on observations
obtained with the Phoenix infrared spectrograph, developed and operated
by the National Optical Astronomy Observatory.}}

\author{Verne V. Smith\footnote{Visiting Astronomer, Gemini South
Observatory}}
\affil{Department of Physics, University of Texas El Paso, El Paso, TX 79968 
USA; verne@barium.physics.utep.edu}

\author{Takashi Tsuji}
\affil{Institute of Astronomy, The University of Tokyo, Mitaka, JP Tokyo 181,
Japan; ttsuji@ioa.s.u-tokyo.ac.jp}

\author{Kenneth H. Hinkle}
\affil{National Optical Astronomy Observatory\footnote{Operated by
Association of Universities for Research in Astronomy, Inc., under
cooperative agreement with the National Science Foundation}, 
P.O. Box 26732, Tucson, AZ 85726 USA; khinkle@noao.edu}

\author{Katia Cunha$^{3}$}
\affil{Observatorio Nacional, Rua General Jose Cristino 77, Sao Cristovao,
20921-400 Rio de Janeiro, Brazil; katia@on.br}

\author{Robert D. Blum}
\affil{Cerro Tololo InterAmerican Observatory$^{3}$, 950 N. Cherry St.,
Tucson, AZ 85719 USA; rblum@ctio.noao.edu}

\author{Jeff A. Valenti}
\affil{Space Telescope Science Institute, 3700 San Martin Dr., Baltimore,
MD 21218 USA; valenti@stsci.edu}

\author{Stephen T. Ridgway, Richard R. Joyce}
\affil{National Optical Astronomy Observatory$^{4}$, P. O. Box 26732,
Tucson, AZ 85726  USA; sridgway@noao.edu, rjoyce@noao.edu}

\author{Peter Bernath}
\affil{Department of Chemistry, University of Waterloo, Waterloo,
ON N2L 3G1, Canada; bernath@uwaterloo.ca}

\bigskip
\author{Submitted to the Astrophysical Journal Letters}

\begin{abstract}

We report on the analysis of high-resolution infrared spectra of
the newly discovered brown dwarf $\epsilon$ Indi B.  This is the
closest known brown dwarf to the solar system, with a distance
of 3.626 pc.  Spectra covering the ranges of
$\lambda$2.308--2.317$\mu$m and $\lambda$1.553--1.559$\mu$m were
observed at a resolution of $\lambda$/$\Delta$$\lambda$=R=50,000.
The physical parameters of effective temperature and surface gravity
are derived for $\epsilon$ Ind Ba by comparison with model
spectra calculated from atmospheres computed using unified cloudy
models.  The results are T$_{\rm eff}$= 1500$\pm$100K, log g= 5.2$\pm$0.3
(in units of cm s$^{-2}$), placing it in the critical boundary between
the late-L and early-T dwarfs.  The high spectral resolution also 
allows us to measure an accurate projected rotational
velocity, with vsin($\iota$)= 28$\pm$3 km s$^{-1}$.
Combined with a published luminosity for $\epsilon$ Ind Ba (with
log(L/L$_{\odot}$)= -4.67), the derived parameters result in a
``spectroscopic'' mass estimate of $\sim$32M$_{\rm Jupiter}$, a
radius of $\sim$ 0.07R$_{\odot}$, and a maximum rotational period
of $\sim$ 3.0 hours.  A compilation and comparison of effective 
temperatures derived from spectroscopy using model atmospheres versus
those derived from luminosities and theoretical M$_{\rm bol}$--radius
relations reveals a systematic disagreement in the T$_{\rm eff}$ scale.
The source of this disagreement is unknown. 

\end{abstract}


\section{INTRODUCTION}

The recently discovered nearby T-dwarf companion to 
$\epsilon$ Indi (Scholz et al. 2003)
will be important in improving our understanding of
the behavior of the brown dwarfs that fall within the newly defined
L and T spectral types (e.g., Kirkpatrick et al. 1999; Geballe et al. 2002).
The T-dwarf reported by Scholz et al. (2003) was discovered by Volk et al.
(2003) to be a close
optical double consisting of an early T dwarf ($\epsilon$ Indi B) and a
late T dwarf ($\epsilon$ Indi C)
separated by 0.6 arcseconds.  
With an accurately known distance of 3.626$\pm$0.009 pc, $\epsilon$
Indi Ba and Bb are the nearest known brown dwarfs.  They share a common proper
motion with the K5V star $\epsilon$ Indi, lying at a projected distance
of $\sim$1460 AU from their presumed primary star.  The age
of $\epsilon$ Indi itself has been estimated by Lachaume et al. (1999)
to be $\sim$0.8 to 2.0 Gyr: this age estimate is based upon its rotational
velocity and Ca II K-line emission.  $\epsilon$ Indi Ba and Bb are thus
brown dwarfs with very well-defined luminosities and approximate ages.

We present the first high-resolution infrared spectroscopic observations of
$\epsilon$ Indi B.  Synthetic spectra computed from unified cloudy
models by Tsuji (2002) are compared to the observed
high-resolution spectra; these comparisons are used to derive
the stellar parameters of effective temperature
(T$_{\rm eff}$), surface gravity (defined as log g), and projected
rotational velocity (vsin(i)).

\section{OBSERVATIONS}

High-resolution infrared (IR) spectra were obtained on $\epsilon$ Indi Ba 
using the 8.1 m Gemini South reflecting telescope and the NOAO
Phoenix spectrometer (Hinkle et al. 1998).  This instrument is a
cryogenically cooled echelle spectrograph that uses order separating
filters to isolate individual echelle orders.  The detector is a
1024x1024 InSb Aladdin II array.  The size of the detector in the
dispersion direction limits the wavelength coverage in a single
exposure to about 0.5 \% (1550 km s$^{-1}$ or $\sim$0.0120$\mu$m at
2.3$\mu$m and $\sim$0.0080$\mu$m at 1.6$\mu$m).  One edge of the
detector is blemished so the wavelength coverage on all but the
brightest source is typically trimmed a few percent to avoid this area.  
The spectra discussed here were observed with the widest (0.35
arcsecond) slit resulting in a spectral resolution of
R=$\lambda$/$\Delta$$\lambda$= 50,000.  Two spectral regions
were observed, with one centered at $\lambda$=2.314$\mu$m and the
other centered at 1.555$\mu$m.  These spectral regions sample
crucial diagnostic lines from the molecules CO, H$_{2}$O, and
CH$_{4}$. 

At the time of our observations we were unaware of the existence of
$\epsilon$ Indi Bb.  Following the discovery (Volk et al. 2003) we
re-examined our acquisition images.  Acquisition images of
$\epsilon$ Indi B at 1.647 $\mu$m taken on 2003 August 13 under
good conditions (0.4" FWHM delivered image quality (DIQ))
confirm the Volk et al. (2003) detection.  The
companion can also be seen in 1.558 $\mu$m images on 2002 December 29,
although it is less well-resolved due to inferior seeing (0.8"
arcsec DIQ).  It is, however, barely perceptible at 2.321 $\mu$m (2003
January 16) with 0.4" DIQ.  The issue relevant to the current
investigation is the extent to which the spectrum of $\epsilon$ Indi B
might suffer contamination from a nearby companion.  The observed image
profile at 1.647 $\mu$m is well fit by a model in which the companion
is fainter by 1.9 mag and is 0.65" away at a position angle of
125 degrees.  At 2.321 $\mu$m the magnitude difference is $>$ 3 between
the two stars, so that the companion could contribute no more than 6\%
to the spectrum in this region.  Our data prevent an estimate at 1.558
$\mu$m but Volk et al. (2003) report a difference of 1.3 magnitudes.
In addition the position angle is at 35 degrees to the 0.35 arcsecond
slit.  Detailed examination of the spectral images showed no trace of
$\epsilon$ Indi Bb at the location expected so we are confident that
our spectra are contributed almost entirely by $\epsilon$ Indi Ba.

Each program star was observed along the slit at two or three 
separate positions separated by 4" to 5" on the sky: the delivered image FWHM 
at the spectrograph varied from 0.25"-0.80" during the nights that spectra
were taken, so stellar images at different positions on the slit were
well separated on the detector. 
Equal integration times were used for a particular program star during
a particular set of observations.  With this observing strategy,
sky and dark backgrounds are removed by subtracting one integration
from another (the star being at different positions on the detector
array).  During each night, 10 flat-field and 10 dark images were
recorded for each given wavelength setting of the echelle.  A hot star,
with no intrinsic spectral lines in the regions observed, was also
observed each night in each observed wavelength region.  

Two-dimensional images were reduced to one-dimensional spectra using
an optimal extraction algorithm described in Johns-Krull, Valenti, \&
Koresco (1999). Wavelengths and telluric corrections in the 2.31 $\mu$m
region were determined by fitting observed telluric features with a
scaled atmospheric transmission function from Wallace \& Hinkle (2001).
There are no significant telluric features in the 1.56 $\mu$m region,
so the wavelength scale was determined by matching a spectrum of HR
1629 (K4 III) with an IR FTS atlas of Arcturus (Hinkle, Wallace, \&
Livingston 1995)

Figure 1 illustrates the combined and reduced spectra for $\epsilon$
Indi Ba in both the 2.313$\mu$m and 1.556$\mu$m regions, with the
wavelengths plotted as air wavelengths.  
The data points have been smoothed
over the slit width of 4-pixels and the final signal-to-noise ratio
is about 30--40.  The 2.31$\mu$m region contains strong 
vibration-rotation lines
from the first overtone bands of CO (here (2--0) lines from
$^{12}$C$^{16}$O), as well as some weak, blended H$_{2}$O features, and 
weak absorption from methane.  Detectable spectral features at
1.55$\mu$m are not as strong as at 2.31$\mu$m, nor as well-defined
as the individual vibration-rotation CO lines, and consist mostly of blended
H$_{2}$O features.  Two of the stronger features can be assigned to mainly 
two H$_{2}$O lines (as marked in Figure 1) on the basis of the study
of H$_{2}$O by Tereszchuk et al. (2002). 

\section{ANALYSIS AND DISCUSSION}

The observed spectra of $\epsilon$ Ind Ba at 2.313$\mu$m and 1.556$\mu$m
are compared to synthetic spectra calculated from model atmospheres
as discussed by Tsuji (2002).  These models are so-called ``unified
cloudy models'' in which dust is allowed to exist in the photosphere
over a limited range defined by a condensation temperature, T$_{\rm cond}$,
and a critical temperature, T$_{\rm cr}$, such that dust is found in
the region of T$_{\rm cr}$ $\le$ T $\le$ T$_{\rm cond}$.  At the critical
temperature, dust grains become so large that they precipitate from the
the photosphere.  The models employed in this analysis are computed with
plane parallel geometry, in hydrostatic equilibrium, and have solar
abundances. 

Figure 2 illustrates a comparison of observed and synthetic spectra,
for both the 2.313$\mu$m (top panel) and 1.556$\mu$m (bottom panel)
regions.  The 2.313$\mu$m spectrum contains 
strong $^{12}$C$^{16}$O (2--0) lines and some weak H$_{2}$O features.
The 1.556$\mu$m spectrum exhibits primarily H$_{2}$O absorption,
with these features composed of many blended individual spectral lines.
In the top panel, the comparison
synthetic spectra span effective temperatures from 1400K
to 1800K and these models have surface gravities of log g= 5.5 (in
units of cm s$^{-2}$).  This particular spectral region is
illustrated as any CH$_{4}$ absorption 
beginning near $\lambda$$\sim$ 2.3158$\mu$m, and clearly apparent in the
models with T$_{\rm eff}$=1400K or 1500K, is very temperature sensitive.
Its observed absence (or extreme weakness) in $\epsilon$ Ind Ba indicates
that T$_{\rm eff}$= 1600K; higher effective temperatures begin to
produce CO lines that are too weak.   The CO lines that dominate this
region are not very sensitive to gravity over the expected range for
the brown dwarfs with these approximate temperatures; however, the
1.556$\mu$m region features dominated by H$_{2}$O are more sensitive
to surface gravity, as shown in the bottom panel of Figure 2.
Here, a slightly lower effective temperature is derived,
with T$_{\rm eff}$=1400K, and the gravity sensitive H$_{2}$O
features indicate that log g$\sim$5.0 to 5.5.  Higher effective temperatures
produce H$_{2}$O absorption features that are too weak for any
reasonable surface gravity, while temperatures much lower than
T$_{\rm eff}$=1400K (say 1300K) produce increasingly strong CH$_{4}$
absorption, which looks nothing like the observed spectrum of $\epsilon$ 
Ind B.  This effect is shown by the T$_{\rm eff}$=1300K model spectrum in
the bottom panel of Figure 2 that is offset vertically from the observed
spectra and 1400K models.  The offset is done as the different absorption
features in the 1300K model (that are caused by increasing CH$_{4}$ and
decreasing H$_{2}$O absorption), if overlaid on the observed spectrum, 
would produce merely confusion. 

Fits to the line
profiles (primarily the strong CO lines), as shown in Figure 2, also
yield the projected rotational velocity, with 
vsin($\iota$)=28 km s$^{-1}$ (with an uncertainty of $\pm$3 km s$^{-1}$). 
Wavelength shifts between observed and synthetic spectra also provide
an accurate radial velocity for $\epsilon$ Ind B, which we find to be
V$_{\rm heliocentric}$= --41.0$\pm$0.7 km s$^{-1}$.  This radial velocity
is very close to the published value for 
$\epsilon$ Ind A's velocity of --39.6$\pm$0.8 km s$^{-1}$ from
Wielen et al. (1999).  The similarity of radial velocities for both
$\epsilon$ Ind A and Ba strengthens their physical association, as
argued by Scholz et al. (2003) based on their respective distances
and proper motions. 

The combination of
both the 2.314$\mu$m and 1.555$\mu$m high-resolution spectra and their 
comparison to synthetic spectra result in values for temperature and
gravity in $\epsilon$ Ind Ba to be T$_{\rm eff}$=1400--1600K and
log g=5.0--5.5.  Taking the average of these values as being the
best estimates we find an effective temperature of 1500K and a gravity
of log g=5.25.  Scholz et al. (2003) used photometry to derive the
luminosity of $\epsilon$ Ind Ba and found it to be
log(L/L$_{\odot}$)=--4.67.  This luminosity can be combined with our
estimates of temperature and gravity to yield a spectroscopic mass
estimate of 32M$_{\rm Jupiter}$.  Scholz et al. (2003) derive a mass
of 40-60M$_{\rm Jupiter}$, which is close to the mass derived
from the high-resolution spectra.  In addition, given the luminosity
and effective temperature, the radius of $\epsilon$ Ind Ba can be
estimated and then combined with the projected rotational velocity
to yield a maximum rotational period.  With log(L/L$_{\odot}$)=--4.67
and T$_{\rm eff}$= 1500K, a radius of (R/R$_{\odot}$)= 0.069 is
derived.  Given vsin($\iota$)=28 km s$^{-1}$, then the maximum
rotational period for $\epsilon$ Ind Ba will be 3.0 hours.

The effective temperature derived by Scholz et al. (2003) is 1260K
and results from the luminosity combined with the distance and a
radius defined by a theoretical M$_{\rm bol}$--radius relation that
results from structural models of brown dwarfs.  This T$_{\rm eff}$
is somewhat lower than our value of 1500K, however
this difference is typical of the differences found to date between
spectroscopically derived effective temperatures and those derived from
structural models.  The differences in T$_{\rm eff}$ also indicate
differences in the implied radii from the two methods.  In our case,
the higher spectroscopic T$_{\rm eff}$ for $\epsilon$ Ind Ba requires
a smaller radius to support its luminosity.  The two techniques are
complementary, in the sense that the radius does not enter into the
computation of the plane parallel atmosphere, but does in the 
M$_{\rm bol}$--radius relation.  Age uncertainties can also affect
derived physical paprameters, especially for the structural models.
Larger sets of comparison measurements
of physical parameters derived from both model atmospheres and structural
models will improve our understanding of the root cause of these
differences.  A comparison of the temperature scales is illustrated in
Figure 3, where spectroscopic T$_{\rm eff}$'s are plotted versus structural
T$_{\rm eff}$'s; the spectroscopic temperatures are taken from
Basri et al. (2000), Leggett et al. (2001), and Schweitzer et al. (2002),
while the structural temperatures are those from Dahn et al. (2002).
The Dahn et al. T$_{\rm eff}$'s also use a M$_{\rm bol}$--radius
relation from structural models in deriving temperatures.  Despite using
different sets of model atmospheres (with differing treatments of dust),
there is a clear trend in the differences between the 
spectroscopically derived effective
temperatures when compared to the structural temperatures from
Dahn et al. (2002).  At higher temperatures, the spectroscopic 
T$_{\rm eff}$'s tend to fall below the structural T$_{\rm eff}$'s,
with the reverse situation at lower temperatures and a crossover point at 
$\sim$1900-2000K.  Our derived T$_{\rm eff}$ for $\epsilon$ Ind B
falls nicely on the cool end of this trend.  We do not speculate here
on the reasons for the systematic differences between structural and
spectroscopic effective temperatures.

The treatment of dust in the photosphere is a crucial ingredient in the
quantitative spectral modelling of the cool L and T dwarfs.  Here we
have used the unified dust models as discussed in detail by Tsuji (2002);
however, we also investigated other dust treatments to see what effects
these would have on the derived physical parameters (primarily 
T$_{\rm eff}$).  Two other sets of model atmospheres were generated: 1)
one in which dust remained in all layers where the thermochemical
conditions allowed dust condensation (case B), and 2) the other being
the case where all dust sank out of the photospheric layers (case C).
These very same effects were also investigated and discussed in detail
by Basri et al. (2000) for a sample of late-M and L dwarfs.  In our case
B, where dust exists over a wide range of depths in the photosphere, the
model absorption lines (CO, H$_{2}$O, and CH$_{4}$) are all much weaker
than the observed absorption lines (the addition of significant dust
opacity over a large region of the photosphere weakens the gas phase
absorption lines), and no realistic fit to the observed spectra is possible 
for any reasonable T$_{\rm eff}$.  For the atmospheres where all dust
sinks from the photosphere (case C), the model molecular absoprtion lines
have more realistic strengths, but the temperature sensitive
CH$_{4}$ appears at even higher effective temperatures; we derive
T$_{\rm eff}$=1800K for $\epsilon$ Indi Ba.  The disagreement with the
structural models is even worse here.  This exercise suggests that some
dust in the photosphere provides a better physical picture in modelling
the high-resolution IR-spectra of $\epsilon$ Ind Ba.  Our initial comparison
here points to the need for more high-resolution spectral analyses across
the temperature range of the L and T dwarfs.

\section{CONCLUSIONS}

We have used high-resolution infrared spectra of the nearest brown
dwarf, $\epsilon$ Ind B, to derive its physical parameters using
comparisons to synthetic spectra calculated from model atmospheres.
The spectroscopic T$_{\rm eff}$=1500K, with log g=5.2, and an
estimated mass of M=32M$_{\rm Jupiter}$.  The projected rotational
velocity is vsin($\iota$)=28 km s$^{-1}$, indicating a maximum
rotational period of $\sim$3.0 hours.  The comparison between the
T$_{\rm eff}$ scales derived from spectroscopic plus model atmosphere
analyses against those derived from M$_{\rm bol}$--radius relations
reveals a significant systematic difference that is unexplained. 

\acknowledgements
The staff of the Gemini South Observatory are to be thanked for 
technical support.  We also thank J. Tennyson for his comments on
the H$_{2}$O assignments, and the referee (G. Basri) for helpful
suggestions in improving this paper.
This research had made use of the SIMBAD database, operated at
CDS, Strasbourg, France.
The work reported here is supported in part by the National Science
Foundation through AST99-87374 (V.V.S.) and NASA through NAG5-9213 (V.V.S.).

\clearpage

\figcaption[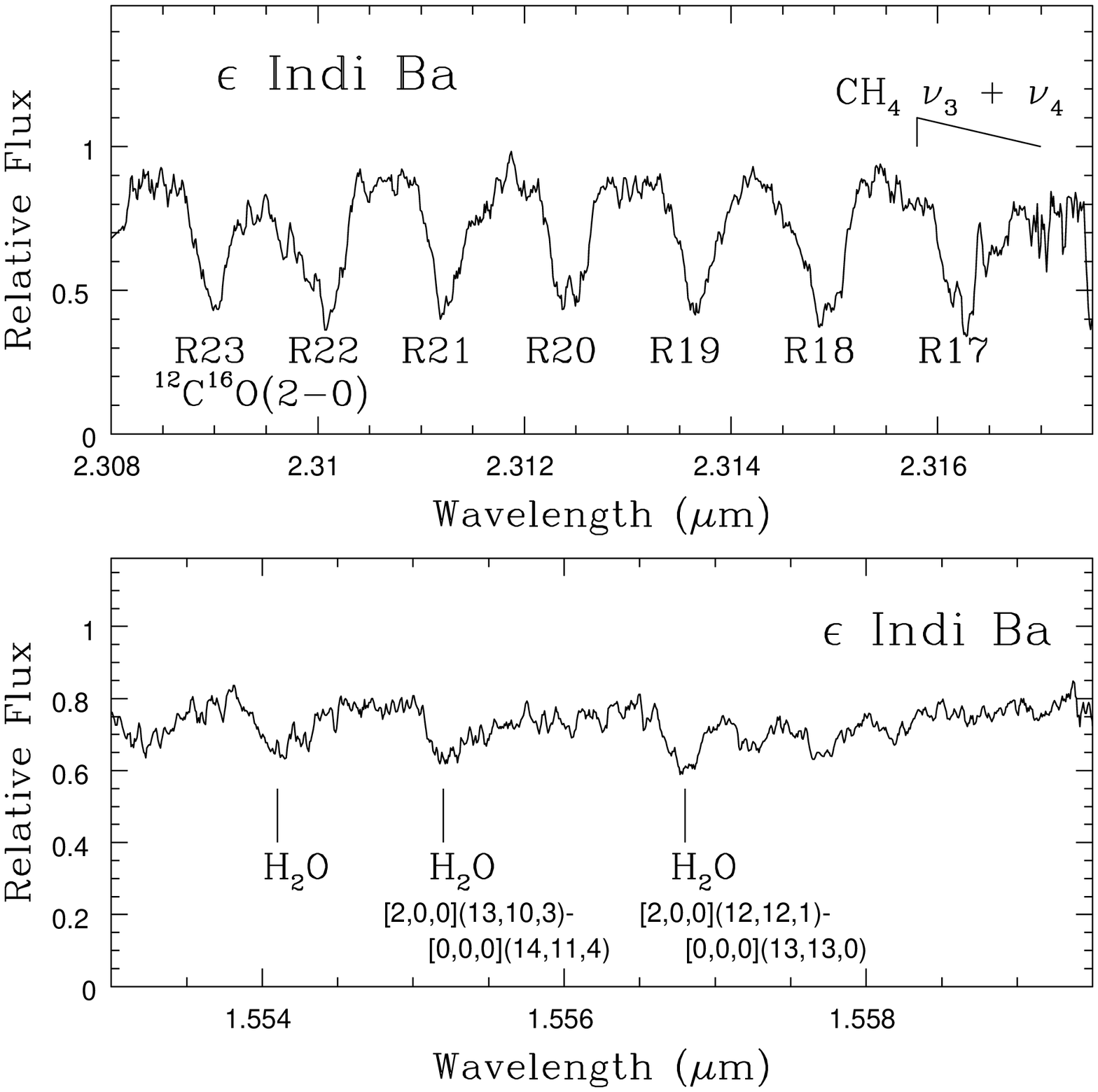]{The $\lambda$2.313$\mu$m (top panel) and
$\lambda$1.556$\mu$m (bottom panel) spectra of $\epsilon$ Indi B,
with the plotted wavelengths being those in air.
The 2.31$\mu$m region is dominated by strong $^{12}$C$^{16}$O(2--0) 
lines that are rotationally broadened. 
The CH$_{4}$ absorption at $\lambda$2.318$\mu$m is
either absent or very weak at this high spectral resolution.
The 1.55$\mu$m region exhibits blended features from H$_{2}$O. 
\label{fig1}}

\figcaption[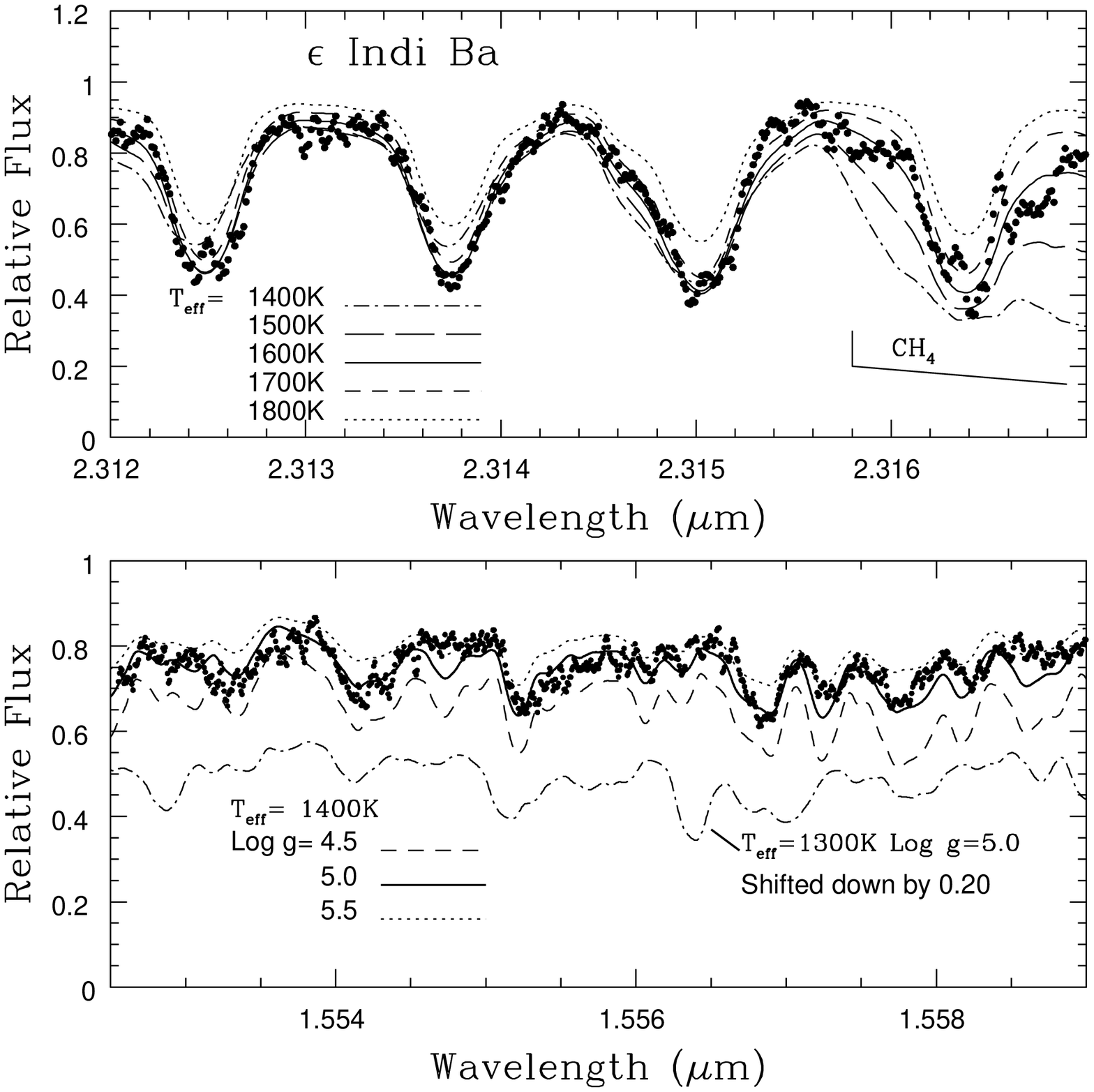]{A comparison of the observed spectra of $\epsilon$
Ind Ba with synthetic spectra that span a range of effective temperatures
and gravities.  In the top panel the $\lambda$2.314$\mu$m region is
shown, with the strong $^{12}$C$^{16}$O lines, as indicated in Figure 1.  
In order to fit the observed line shapes the model spectra must be
broadened by a rotational profile with v$_{\rm rot}$=28 km s$^{-1}$.
This spectral region is illustrated because of the large temperature
sensitivity of the CH$_{4}$ absorption (as well as the CO).  A good fit
is obtained for T$_{\rm eff}$=1600K.  These particular lines are not
very sensitive to gravity over the expected values, so a single-gravity
set of models is shown (with log g= 5.5).  The bottom panel shows the
observed and model spectra comparisons for a single T$_{\rm eff}$= 1400K
but a range of gravities.  Most of the blended spectral features visible
in this wavelength region are from H$_{2}$O, and the relative depth
of the absorption is sensitive to gravity, as illustrated.  Surface
gravities in the range of log g= 5.0 to 5.5 are the overall best fits.
Note that higher effective temperatures will produce extremely weak
H$_{2}$O absorption, while lower T$_{\rm eff}$'s result in increasingly
dominant CH$_{4}$ absorption, which is not observed.  The vertically shifted
model spectrum is for T$_{\rm eff}$=1300K and illustrates the increasingly
different absorption as the temperature decreases (note the strong
feature at 1.5564$\mu$m from CH$_{4}$. 
\label{fig2}}

\figcaption[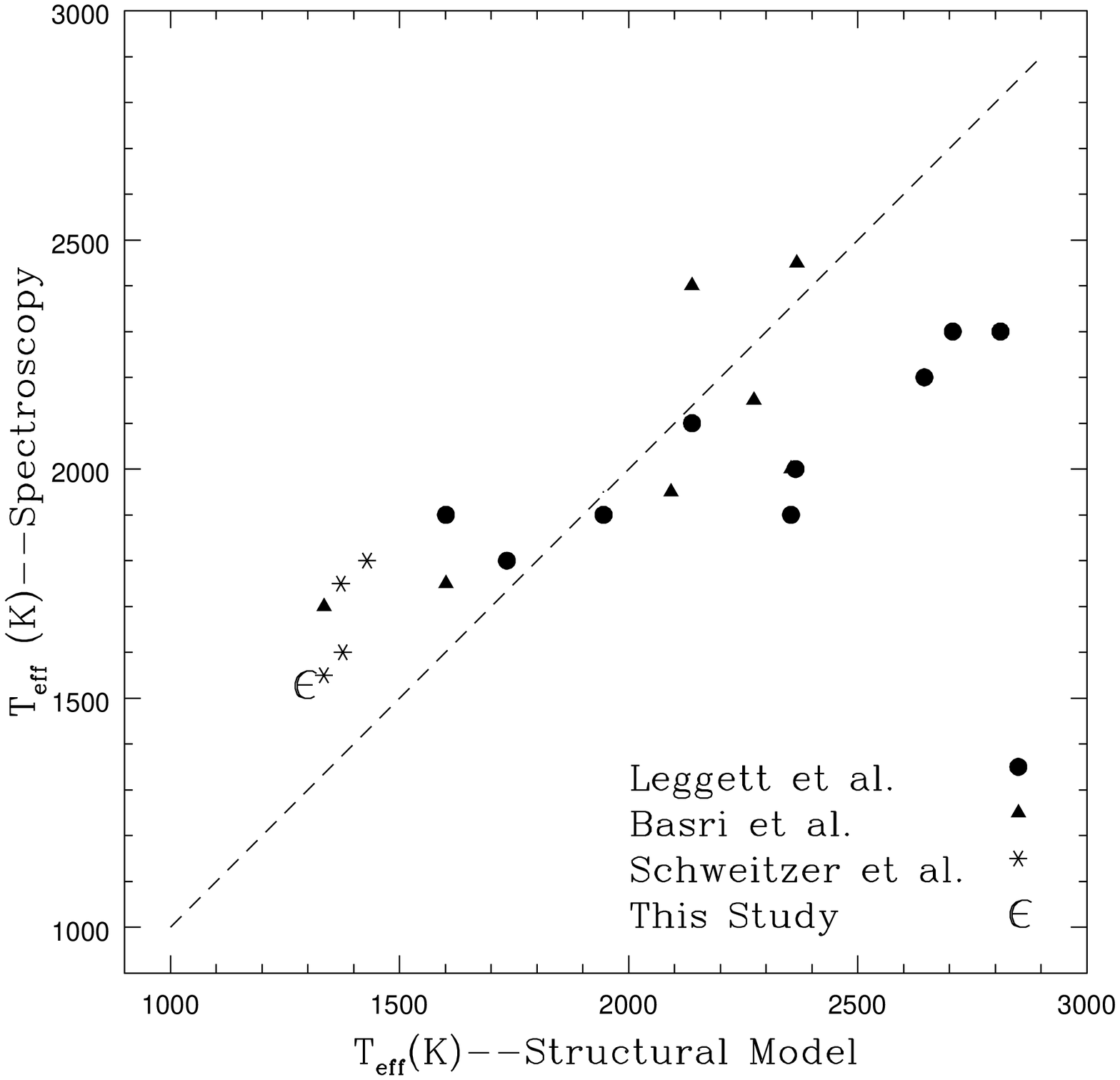]{A comparison of effective temperatures derived from
model atmosphere spectrum synthesis techniques to temperatures derived from
structural models for L and T dwarfs: the structural T$_{\rm eff}$'s are
taken from Dahn et al. (2002), while the spectroscopic temperatures
are from Basri et al. (2000), Leggett et al. (2001), and Schweitzer
et al. (2002).  There is a well-defined systematic trend in the
differences between spectroscopic and structural effective temperatures,
with our result for $\epsilon$ Ind Ba falling at the cool end of this
trend.
\label{fig3}}

\end{document}